\documentclass[11pt,reviewcopy]{elsart}

\usepackage{endfloat}
\usepackage{amssymb}
\usepackage{epsfig}
\usepackage{jtb}

\begin{document}

\begin{frontmatter}

\title{The Effect of a Random Drift on Mixed and Pure Strategies in the Snowdrift Game}
\author{Andr\'e C. R. Martins and Renato Vicente\corauthref{cor1}}
\corauth[cor1]{Email: rvicente@usp.br Fax: +55-11-30916748 Phone: +55-11-30918139}
\address{GRIFE, Escola de Artes, Ci\^encias e Humanidades, Universidade de S\~ao Paulo, Av. Arlindo B\'etio 1000, 03828-000, S\~ao Paulo-SP, Brazil.}

\begin{abstract}
The replicator dynamics of players choosing  either mixed or  pure strategies are usually regarded as equivalent, as long as strategies are played with identical frequencies. In this paper we show  that a population of pure strategies  can be invaded by  mixed strategies in any two-player game with equilibrium coexistence upon the addition of an arbitrarily small amount of noise in the replication process.  
\end{abstract}


\end{frontmatter}

The understanding of the spontaneous emergence of cooperation is  central to uncover the origin of  complex social organizations in populations of selfish genes. The general framework employed \cite{NowakBook} supposes a population  interacting in pairs by playing evolutionary games representing social dilemmas. The individuals choose among a repertoire of behaviors and the outcome of these interactions affects their reproductive fitness. In the simplest case these individuals can be regarded as players of a game with two strategies represented by a $2\times 2$ payoff matrix. As the game is interactively played, individuals with larger payoffs will thrive with larger reproductive fitnesses while those with worse performances will eventually perish.  

Recent experimental evidence \cite{Kumm07} suggests that the iterated Snowdrift Dilemma (SD) \cite{Hauert05} may be more suitable for explaining  the high levels of cooperation observed among unrelated human individuals than the more intensively studied  iterated Prisoners' Dilemma.   In a SD each player can in each turn either cooperate (C) or defect (D). The cooperative behavior by at least one individual yields a benefit $b$ to both players. If both individuals cooperate each one contributes  $c/2$  to the total cost, otherwise the total cost $c$ is  payed by the cooperator alone. Mutual defection yields no benefit or cost to any player. 

The replicator dynamics \cite{SigmundBook} for the evolutionary SD game yields a simple prediction. If there are only cooperators in the population, a defector will have the advantage and will invade. Similarly, in a population of defectors, a cooperator can also invade. A fixed point is reached when cooperators are observed with probability $p_C=1-c/(2b-c)$. 

It is generally  assumed  that it makes no difference whether the population is composed by genes playing pure defective (D) or cooperative (C) strategies in a proportion defined by $p_C$ or by genes playing mixed strategies with probability of cooperating  $p_C$. Notice that a gene can play a mixed strategy either by creating individuals who play mixed strategies or by having a proportion of its carriers to behave as cooperators while the others behave as defectors. In other words an arbitrary  population of pure strategists is thought to be stable against a population of mixed strategy players, given that the probability of cooperating is tunned to be $p_C$.

In the following we will analyze the effect of a random drift on pure and mixed strategies playing games with equilibrium coexistence having the SD as a major example.

\begin{figure}
\begin{center}
\vglue -4.mm
\psfig{file=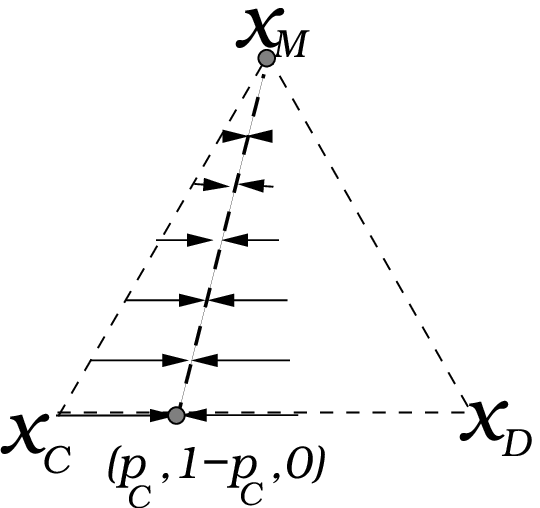,width=5cm}
\caption{Simplex for a game with equilibrium coexistence, pure and, mixed strategies. The thick dashed line represents fixed points. Stable eigen-directions are represented by arrows with lengths proportional to the related eigenvalues.}
\label{fig2}
\vglue -2.mm
\end{center}
\end{figure}

Let us suppose a well-mixed population playing an evolutionary two-player game with strategies $C$ and $D$ and payoff matrix given by:
\begin{equation}
A=\left(\begin{array}{cc}
	a_{CC}& a_{CD}\\
	a_{DC}& a_{DD}
\end{array}\right),
\end{equation}
with $a_{CC}<a_{DC}$ and  $a_{CD}>a_{DD}$, implying strategies that coexist at equilibrium with cooperators at frequency:
\begin{equation}
\label{C_proportion}
p_C=1-\frac{a_{CD}-a_{DD}}{a_{DD}-a_{CC}+a_{CD}+a_{DC}}.
\end{equation}

Suppose this population to be  composed  of  mixed  strategists at frequency $x_M$ playing $C$ with probability $p_C$, pure defectors and pure cooperators at  frequencies $x_D$ and $x_C$, respectively.

The replicator dynamics is given by \cite{NowakBook}:
\begin{equation}
\dot{x}_a=x_a\left(f_a -\bar{f}\right),
\label{replicator_3eqs}
\end{equation}
where $x_a$ and $f_a$ are, respectively, the frequency and  the payoff (or fitness) of type $a$ players and $\bar{f}=\sum_a x_af_a$ is their mean fitness. Considering that $x_M + x_C + x_D =1$, we can eliminate one of the three equations in (\ref{replicator_3eqs}) to write:
\begin{eqnarray}
\dot{x}_C&=&(f_C-f_D)\left[(1-p_C) (x_C-x_C^2) + p_C x_Cx_D\right]\\
\dot{x}_D&=&(f_C-f_D)\left[p_C (x_D^2-x_D)-(1 - p_C) x_Cx_D\right]\nonumber,
\label{replicator_2eqs}
\end{eqnarray}
with
\begin{equation}
f_C-f_D=-\frac{a_{CD}-a_{DD}}{1-p_C}\left[(1-p_C) x_C-p_C x_D\right].
\end{equation}

It can be easily verified that  the dynamical system in (\ref{replicator_2eqs}) has a line
of fixed points defined by $x_D= x_C(1-p_C)/p_C $. The linear stability of these fixed points  can be studied by calculating the  Jacobian matrix in the neighborhood of the line to find:
\begin{equation}
{\cal J}=-(a_{CD}-a_{DD})x_C\left(\begin{array}{cc}
	(1-p_C)& -p_C\\
	-(1-p_C)& p_C
\end{array}\right),
\end{equation}
which has a stable eigen-direction $v_1=(1,-1)$  with eigenvalue $\lambda_1=-(a_{CD}-a_{DD})x_C$ and a degenerate eigen-direction $v_0=(1,(1-p_C)/p_C)$. As expected, this analysis reveals that a population composed by pure strategies at equilibrium and by mixed strategies is stable regardless of the frequency $x_M$ of mixed strategists. 

In figure \ref{fig2} we represent this general picture in a simplex with special attention to the fact that stable eigenvalues  depend on $x_C$. 

\begin{figure}
\begin{center}
\vglue -4.mm
\psfig{file=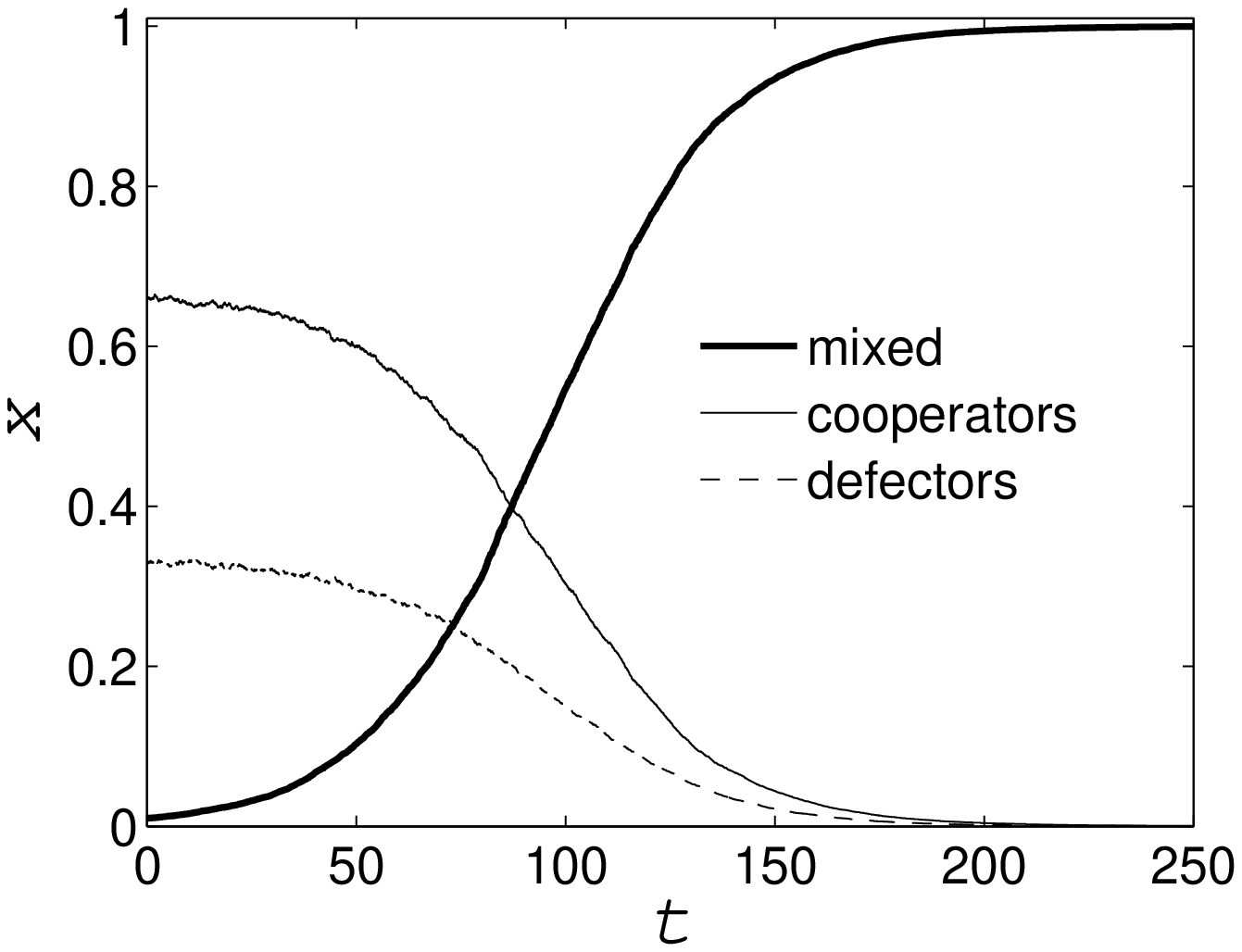,width=9cm}
\caption{Simulation of the Snowdrift game with $b=4$ and $c=2$ starting with a population of pure strategy playing at the evolutionary equilibrium and one single mixed strategy.}
\label{fig:sim}
\vglue -2.mm
\end{center}
\end{figure}

To  assess the behavior along the stable line we introduce a random perturbation as $x_D=x(1-p_C)/p_C+ \epsilon$, $x_C=x$ and $x_M=1-x_C-x_D-\epsilon$ with $\epsilon>0$ if $x=0$ and $\epsilon<0$ if $x=p_C$ to find:
\begin{equation}
\dot{x}\simeq p_C (a_{CD}-a_{DD})\epsilon x,
\label{perturb}
\end{equation}
Let us now suppose that the replication of each and every one of the strategies is randomly influenced by global environmental changes. This may be represented by adding noise in the  mapping of payoffs into fitnesses. To assess this scenario we replace the perturbation in (\ref{perturb}) by a Wiener process as $\epsilon_t dt=\sigma dW_t$ \cite{OksendalBook}.  We can then show that this component leads to  a drift toward a population composed exclusively  by mixed strategies (point $x_M$ in figure \ref{fig2}). That can be seen by changing variables in (\ref{perturb}) to write the following stochastic differential equation in It\^o form:
\begin{equation}
d\log(x)=\alpha dW_t -\frac{1}{2}\alpha^2dt,
\label{stochastic}
\end{equation}   
where $\alpha= \sigma p_C (a_{CD}-a_{DD})$. We now consider an initial  population with  pure strategies and a positive, but very small, frequency of mixed strategies $x\simeq p_C$. Assuming very small fluctuations ($\sigma$  very small), equation (\ref{stochastic}) yields the following approximation for the  probability density for $x$:
\begin{equation}
\rho_t(x)=\frac{1}{\alpha\sqrt{2\pi t}} \exp\left[-\frac{\left(\log{\frac{x}{p_C}}+\frac{1}{2}\alpha^2 t\right)^2}{2\alpha^2 t}\right],
\end{equation}   
that converges as $t\rightarrow\infty$ to  $\rho_\infty(x)=\delta(x)$, to say, to populations concentrated around  $x_M=1$. In figure \ref{fig:sim} we show a simulation for the replicator dynamics of the SD game with $b=4$ and $c=2$. The initial condition is set to be the evolutionary equilibrium with cooperators to be found at frequency $p_C$ and a single mixed strategy with probability $p_C$ of cooperating. As predicted, the population of pure strategies is led to extinction  by mixed strategies.

The general picture that emerges can be described as follows. Suppose that the population has been brought to a rest point. From there, any change in the proportion of the mixed strategies that keeps the proportion of the pure strategies unaltered will not change the matching probabilities when the game is played and therefore, this will correspond to a perturbation along a stable eigen-direction, with no consequences other than the random change.

A different scenario is observed if the ratio of the pure strategies is somehow altered. When that happens, matching probabilities will change to a point near the rest point. And, while the mixed strategy is still playing the evolutionary equilibrium answer to the game, the pure strategies are no longer doing that. This means that some of them might be better than average (those that have become rarer), while others will certainly be worse (the strategies that became more common). This will cause the system to move closer to the rest point, restoring the equilibrium, as expected in the basin of attraction. However, an important effect happens here. The mixed strategy was playing the best answer and therefore, its fitness becomes larger than the average until the equilibrium is restored, regardless of the direction of departure. This means that any time a drift that changes the relative proportions of pure strategies happens, the mixed strategy will increase its proportion. In the long run, it will slowly dominate the game and the drift will cause the extinction of all pure strategies.
 
We emphasize that the conclusion above holds for any game with equilibrium coexistence and we have used the SD case as an example. That means that any analysis of evolutionary processes or of the emergence of cooperation should  also be performed without the supposition that strategies breed true. In the long run, mixed strategies will have a higher probability of survival and this fact should be taken into consideration when evolutionary dynamics is considered.


\begin{thebibliography}{8}
\expandafter\ifx\csname natexlab\endcsname\relax\def\natexlab#1{#1}\fi
\expandafter\ifx\csname url\endcsname\relax
  \def\url#1{\texttt{#1}}\fi
\expandafter\ifx\csname urlprefix\endcsname\relax\def\urlprefix{URL }\fi
\providecommand{\selectlanguage}[1]{\relax}

\bibitem[{Doebeli \& Hauert(2005)}]{Hauert05}
\textsc{Doebeli, M. \& Hauert, C.} (2005).
\newblock Models of cooperation based on the prisoner's dilemma and the
  snowdrift game.
\newblock \emph{Ecology Letters} \textbf{8}, 748--766.

\bibitem[{Hofbauer \& Sigmund(1998)}]{SigmundBook}
\textsc{Hofbauer, J. \& Sigmund, K.} (1998).
\newblock \emph{Evolutionary Games and Polpulation Dynamics}.
\newblock Cambridge: Cambridge University Press.

\bibitem[{Kummerli {\it et al.}(2007)}]{Kumm07}
\textsc{Kummerli, R. et al.} (2007).
\newblock Human cooperation in social dilemmas: comparing the snowdrift game
  with the prisoner's dilemma.
\newblock \emph{Proc. R. Soc. B} \textbf{274}, 2965--2970.

\bibitem[{Nowak(2006)}]{NowakBook}
\textsc{Nowak, M.} (2006).
\newblock \emph{Evolutionary Dynamics: Exploring the Equations of Life}.
\newblock Boston: Belknap Press.

\bibitem[{Oksendal(2000)}]{OksendalBook}
\textsc{Oksendal, B.} (2000).
\newblock \emph{Stochastic Differential Equations}.
\newblock Springer-Verlag.

\end{thebibliography}
\end{document}